\documentclass[a4paper,12pt, epsfig]{article}
\usepackage{epsfig}
\usepackage{amssymb}
\usepackage{amsfonts}
\usepackage{amsmath}

\newskip\humongous \humongous=0pt plus 1000pt minus 1000pt

\newif\ifdtup

\catcode`\@=11

\@addtoreset{equation}{section}
\def\theequation{\thesection.\arabic{equation}}

\def\@normalsize{\@setsize\normalsize{15pt}\xiipt\@xiipt
\abovedisplayskip 14pt plus3pt minus3pt%
\belowdisplayskip \abovedisplayskip
\abovedisplayshortskip \z@ plus3pt%
\belowdisplayshortskip 7pt plus3.5pt minus0pt}

\def\small{\@setsize\small{13.6pt}\xipt\@xipt
\abovedisplayskip 13pt plus3pt minus3pt%
\belowdisplayskip \abovedisplayskip
\abovedisplayshortskip \z@ plus3pt%
\belowdisplayshortskip 7pt plus3.5pt minus0pt
\def\@listi{\parsep 4.5pt plus 2pt minus 1pt
     \itemsep \parsep
     \topsep 9pt plus 3pt minus 3pt}}

\relax

\catcode`@=12

\catcode`\@=11

\def\section{\@startsection{section}{1}{\z@}{3.5ex plus 1ex minus
   .2ex}{2.3ex plus .2ex}{\large\bf}}

\def\thesection{\arabic{section}}
\def\thesubsection{\arabic{section}.\arabic{subsection}}

\def\appendix{\setcounter{section}{0}
 \def\thesection{Appendix \Alph{section}}
 \def\thesubsection{\Alph{section}.\arabic{subsection}}
 \def\theequation{\Alph{section}.\arabic{equation}}}

\def\SymBoxes#1#2#3#4{\newdimen\un@t \un@t#3%
\raisebox{#1}{\rule{#2\un@t}{#4}\hskip-#2\un@t
\@tempdimb\un@t \advance\@tempdimb by-#4\@tempcntb#2\relax%
\@whilenum{\@tempcntb>0}\do{
\rule{#4}{\un@t}\hskip\@tempdimb \advance\@tempcntb by\m@ne}%
\hskip-#2\un@t \rule[\un@t]{#2\un@t}{#4}%
\rule[\un@t]{#4}{#4}\hskip-#4
\rule{#4}{\un@t}}\hskip-#4}                

\begin{document}

\newcommand{\beq}{\begin{equation}}
\newcommand{\eeq}{\end{equation}}
\newcommand{\bea}{\begin{eqnarray}}
\newcommand{\eea}{\end{eqnarray}}
\newcommand{\beas}{\begin{eqnarray*}}
\newcommand{\eeas}{\end{eqnarray*}}
\newcommand{\defi}{\stackrel{\rm def}{=}}
\newcommand{\non}{\nonumber}
\newcommand{\bquo}{\begin{quote}}
\newcommand{\enqu}{\end{quote}}
\renewcommand{\(}{\begin{equation}}
\renewcommand{\)}{\end{equation}}
\def \eqn#1#2{\begin{equation}#2\label{#1}\end{equation}}
\def\de{\partial}
\def\Tr{ \hbox{\rm Tr}}
\def\H{ \hbox{\rm H}}
\def\HE{ \hbox{$\rm H^{even}$}}
\def\HO{ \hbox{$\rm H^{odd}$}}
\def\K{ \hbox{\rm K}}
\def\Im{ \hbox{\rm Im}}
\def\Ker{ \hbox{\rm Ker}}
\def\const{\hbox {\rm const.}}
\def\o{\over}
\def\im{\hbox{\rm Im}}
\def\re{\hbox{\rm Re}}
\def\bra{\langle}\def\ket{\rangle}
\def\Arg{\hbox {\rm Arg}}
\def\Re{\hbox {\rm Re}}
\def\Im{\hbox {\rm Im}}
\def\exo{\hbox {\rm exp}}
\def\diag{\hbox{\rm diag}}
\def\longvert{{\rule[-2mm]{0.1mm}{7mm}}\,}
\def\a{\alpha}
\def\dag{{}^{\dagger}}
\def\tq{{\widetilde q}}
\def\p{{}^{\prime}}
\def\W{W}
\def\N{{\cal N}}
\def\hsp{,\hspace{.7cm}}
\newcommand{\C}{\ensuremath{\mathbb C}}
\newcommand{\Z}{\ensuremath{\mathbb Z}}
\newcommand{\R}{\ensuremath{\mathbb R}}
\newcommand{\rp}{\ensuremath{\mathbb {RP}}}
\newcommand{\cp}{\ensuremath{\mathbb {CP}}}
\newcommand{\vac}{\ensuremath{|0\rangle}}
\newcommand{\vact}{\ensuremath{|00\rangle}}
\newcommand{\oc}{\ensuremath{\overline{c}}}
\begin{titlepage}
\begin{flushright}
\end{flushright}
\bigskip
\def\thefootnote{\fnsymbol{footnote}}

\begin{center}
{\Large {\bf
The Black Di-Ring:}} \\ {\large
{\bf An Inverse Scattering Construction}
}
\end{center}

\bigskip
\begin{center}
{\large  Jarah EVSLIN\footnote{\texttt{jevslin@ulb.ac.be}} and Chethan
KRISHNAN\footnote{\texttt{Chethan.Krishnan@ulb.ac.be}}}\\
\end{center}

\renewcommand{\thefootnote}{\arabic{footnote}}

\begin{center}
\vspace{1em}
{\em  { International Solvay Institutes,\\
Physique Th\'eorique et Math\'ematique,\\
ULB C.P. 231, Universit\'e Libre
de Bruxelles, \\ B-1050, Bruxelles, Belgium\\}}

\end{center}

\noindent
\begin{center} {\bf Abstract} \end{center}
We use the inverse scattering method (ISM)
to derive concentric
non-supersymmetric black rings.
The approach used here is fully five-dimensional, and has the modest
advantage that
it generalizes
readily to the construction of more general axi-symmetric
solutions.

\vspace{1.6 cm}

\vfill

\end{titlepage}
\bigskip

\hfill{}
\bigskip

\tableofcontents

\setcounter{footnote}{0}
\section{\bf Introduction}

\noindent
The black ring of Emparan and Reall \cite{Emparan:2001wn}
was the first concrete piece of evidence that in higher
dimensional gravity, the no-hair theorems of 3+1
dimensions need not apply. Their construction explicitly demonstrated that
in an asymptotically flat spacetime with a given ADM mass and angular
momentum, the geometry need not necessarily be that of the Myers-Perry
black hole \cite{Myers:1986un}.

Emboldened by that discovery, a lot of recent work has been
directed towards exploring black rings and related ideas
\cite{Elvang:2004rt}.
One upshot of these investigations is that now we know that there is a
continuous non-uniqueness for black hole solutions in higher dimensions.
Concentric black rings (and the 
black Saturn 
 \cite{Elvang:2007rd, Yazadjiev:2007cd}) are an explicit way
to
realize this degeneracy,
the idea being that you can distribute the angular momenta and the masses
between the two black rings in a continuous way, while still keeping
their total asymptotic values fixed.

Concentric supersymmetric black rings were first constructed in
\cite{Gauntlett:2004wh}, and the restriction to
supersymmetry was lifted
in the work of \cite{Iguchi:2007is}. The technique used in the latter
relies on the clever observation that the problem can essentially be
reduced to four dimensions, and then applying the formalism of
\cite{Manko}. A disadvantage of the lack of a genuinely five
dimensional derivation
is that there is no immediate route that one can pursue in order to
generalize this solution. For instance, to try to add more generic
spins to the solution, or to generalize the construction to more
generic Saturn-like solutions, we would have to tackle the genuinely
five-dimensional problem.
One purpose of this paper is to give a derivation of concentric rings
based on
the general formalism of the inverse scattering method, which does not
rely on the reduction to four dimensions.
The inverse scattering approach that we use here was first used in
the context of higher dimensional gravity in
\cite{Pomeransky:2005sj} and then
further explored in various contexts in
\cite{Tomizawa:2005wv, Senkov}.

The format of this paper is as follows. In the next section we review the
inverse scattering method and the use of Lax pairs for generating new
solutions. Section 3 applies this formalism to the construction of
multiple rings. Once the solution is at hand, we need to impose asymptotic
flatness and the absence of certain singularities. 
These
put some relations between the parameters in the solution.
We conclude with some discussions and possible directions for future
research.

\section{\bf The Inverse Scattering Method: Lax Pairs and Solitons}

In this section we review the inverse scattering method as applied to the
construction of axially symmetric vacuum solutions of Einstein's
equations. The formalism was
developed in four dimensions by Belinski and Zakharov
\cite{Belinsky:1971nt}, a standard textbook is \cite{Belinski:2001ph}.
We will follow the presentation of the method as given in
\cite{Pomeransky:2005sj}, for five dimensions.

In 5D, axial symmetry implies the existence of three
commuting Killing vector fields.  The generic metric with these
assumptions can be written as \cite{Emparan:2001wk, Harmark:2004rm},
\eqn{metric}{ds^2=G_{ab}(\rho,z)dx^adx^b+f(\rho,z)(d\rho^2+dz^2)}
where $a, b=1,2,3$,
and we are free to choose
\eqn{det}{{\rm det}G=-\rho^2.}
If we define two matrices
\eqn{UV}{U\equiv\rho(\partial_\rho G)G^{-1}, \ \ V\equiv\rho(\partial_z
G)G^{-1},}
then Einstein's equations take the form
\begin{eqnarray}
\label{number2} \partial_\rho U+\partial_z V=0, \hspace{0.5in} \\
\partial_\rho (\log f)=-\frac{1}{\rho}+\frac{1}{4\rho}{\rm Tr}(U^2-V^2),
\\
\partial_z(\log f)=\frac{1}{2\rho}{\rm Tr}(UV). \hspace{0.35in}
\end{eqnarray}
The last two equations can be consistently integrated because the first
equation is an integrability condition for them. So the problem is fully
solved, once we fix $G_{ab}$.

The inverse scattering method hinges on the fact that the equations that
need to be solved, namely equations (\ref{det}) and (\ref{number2}), can
be thought of as the compatibility conditions for the following
over-determined set of differential equations:
\eqn{lax}{D_\rho \Psi=\frac{\rho U+\lambda V}{\lambda^2+\rho^2}\Psi, \
D_z \Psi=\frac{\rho V- \lambda U}{\lambda^2+\rho^2}\Psi,}
where
\eqn{covar-der}{D_\rho\equiv\partial_\rho+\frac{2\lambda\rho}{\lambda^2+\rho^2}\partial_\lambda,
\
D_z\equiv\partial_z-\frac{2\lambda^2}{\lambda^2+\rho^2}\partial_\lambda .}
These equations comprise the Lax pair, $\lambda$ is called the spectral
parameter, and the generating matrix $\Psi$ is such that
$\Psi(\lambda=0,\rho,z)=G(\rho,z)$. The first step in the construction
of new solutions, is to start with a seed solution $G_0$, and then
find the  generating matrix $\Psi_0$ that solves (\ref{lax}), with the
appropriate $U_0$ and $V_0$. The generating matrix should satisfy the
condition that $\Psi_0(\lambda=0,\rho,z)=G_0(\rho,z)$. Now, we seek a new
solution of the Lax pair in the form
$\Psi=\chi \Psi_0$ where $\chi$ is called the dressing matrix. Once the
dressing matrix is known, the new solution will be determined as
$G(\rho,z)= \Psi(\lambda=0,\rho,z)$.

We will be interested in finding dressing matrices that satisfy the
ansatz,
\eqn{dress-ansatz}{\chi=1+\sum_k\frac{R_k}{\lambda-\tilde\mu_k},}
where $k$ runs over $1,..,n$: we say that we have an $n$-soliton dressing
matrix. By imposing
conditions on the analyticity structure of the poles in the
$\lambda$-plane, it turns out that we can fix the $\tilde\mu$ to be
\eqn{mu}{\tilde\mu_k=\pm\sqrt{\rho^2+(z-a_k)^2} - (z-a_k),}
where $a_k$ are real constants. We will refer to the positive sign pole as a
soliton $\mu_k$, and the negative sign pole as an anti-soliton
$\bar\mu_k$. In addition to the $a_k$, we also need to specify the $R_k$
(which are {\em not} constants)
in order to fully specify the dressing matrix. It turns out, after some
computation (we refer the interested reader to \cite{Belinski:2001ph} for
details), that this can
be done by specifying $n$ constant vectors with components $m_{0a}^{(k)}$.
These are called the Belinski-Zakharov vectors, and they have
3-components, as implied by the index $a$.
Instead of writing down the
$R_k$ in terms
of $m_{0}^{(k)}$, we will omit the intermediate steps and present the
final solution (the metric $G$) after the $n$-soliton transformation. To
do this, we first
define new vectors $m^{(k)}$:
\eqn{mvector}{m_a^{(k)}=m_{0b}^{(k)}[\Psi_0^{-1}(\lambda=\tilde\mu_k,
\rho,z)]_{ba},}
and the matrix $\Gamma$:
\eqn{gamma}{\Gamma_{kl}=\frac{m_a^{(k)}(G_0)_{ab}m_b^{(l)}}{\rho^2+\tilde\mu_k
\tilde\mu_l}.}
In terms of these, the final metric will be written as
\eqn{finalmetric}{G_{ab}=(G_0)_{ab}-\sum_{kl}\frac{(G_0)_{ac}
m_c^{(k)}(\Gamma^{-1})_{kl}
m_d^{(l)}(G_0)_{db}
}{\tilde\mu_k
\tilde\mu_l}.}
Matrix multiplication along the $a, b,..$-indices is assumed.

The solution as written down in (\ref{finalmetric}) does not always give
rise to the appropriate normalization (\ref{det}) for the final solution.
Instead, for the above $n$-soliton transformation and choice of BZ
vectors, one finds (see equation (8.27) in \cite{Belinski:2001ph})
\eqn{detG}{{\rm
det}G=(-1)^{n}\rho^{2n}\Big(\prod_{k=1}^{n}\tilde\mu_k^{-2}\Big){\rm
det}G_0.}
One way to overcome this difficulty is to only look at
transformations which are of the following two-step form:

Step1. $Subtract$ solitons with trivial BZ-vectors. Trivial, in this
context, means
that the BZ vectors do not mix
components of the diagonal seed metric that we start with.

Step2. $Add$ the same solitons back in
the second step, but this time with non-trivial BZ-vectors.

The reason why this works, is because the BZ-vectors
do not contribute to (\ref{detG}), only the solitons do. And the effect
of the solitons in the first step is annulled by the second step, leaving
us with ${\rm det}G={\rm det}G_0=-\rho^2$. The conformal factor $f$
associated with the final metric can be written as
\eqn{conformal}{f=f_0\frac{{\rm det} (\Gamma_{kl})}{{\rm det}
(\Gamma_{kl}^{(0)})},}
where $\Gamma_{kl}^{(0)}$ is obtained by ``trivializing"
$\Gamma_{kl}$, i.e., by
setting the parameters that make the BZ-vector non-trivial, to zero,
in (\ref{gamma}). The formalism presented here will become more
transparent when we explicitly construct the solution in the following
section.

\section{\bf The Black Di-Ring}

\subsection{Seed Solution and Solitonic Transformations}

As described in the last section, the inverse scattering method uses
certain multi-soliton transformations to generate new solutions of
Einstein's equations from old. So the trick is essentially to guess a
seed solution, the BZ vectors, and the solitons: the formalism
will then churn out the final solution.

A powerful way to handle stationary, axi-symmetric solutions
was invented by Harmark \cite{Harmark:2004rm}
generalizing
earlier work in four and higher
\cite{Emparan:2001wk} dimensions. The idea is that such a spacetime can
be described by certain ``rod configurations". We can describe our
solutions
(both the seed and the
final versions) using these rods. The seed solution for the black
di-ring we take in the form given in figure 1. The construction can be
extended straightforwardly\footnote{In principle. The computational effort
required to derive the final metric grows quickly as we increase the
number of rings.} to more rings by adding the same structure to the left.

\begin{figure}
\begin{center}
\includegraphics[width=0.9\textwidth]{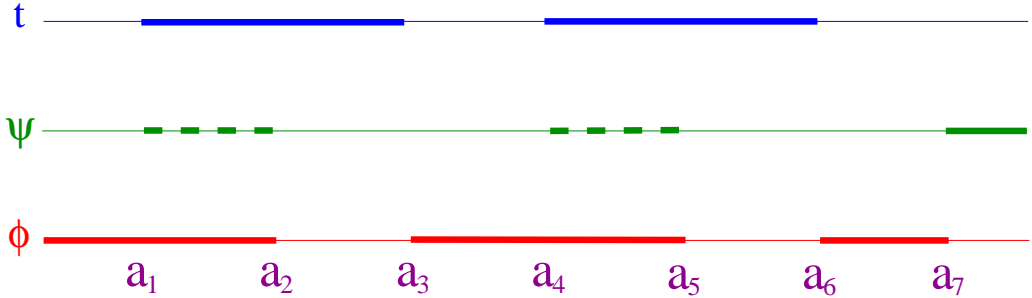}
\caption{Rod configuration for the seed solution}
\label{seedrod}
\end{center}
\end{figure}

Using the standard techniques of \cite{Harmark:2004rm}, we can read off
the seed metric from the seed rod configuration:
\eqn{ringseed}{G_0={\rm
diag}\Big\{-\frac{\mu_1\mu_4}{\mu_3\mu_6},\frac{\rho^2\mu_3\mu_6}
{\mu_2\mu_5\mu_7}, \frac{\mu_2\mu_5\mu_7}{\mu_1\mu_4}\Big\}.}
The elements are the $tt$, $\phi\phi$, $\psi\psi$ components
respectively.
Clearly, this satisfies the normalization condition (\ref{det}).
To complete the description, we write down the conformal factor as well:
\eqn{ringf}{f_0=\frac{k^2 \ \mu_2 \ \mu_5 \ \mu_7\
{\cal R}_{12}{\cal R}_{13}{\cal R}_{15}{\cal R}_{16}{\cal
R}_{17}{\cal
R}_{23}{\cal R}_{24}{\cal
R}_{26}{\cal R}_{34}{\cal R}_{35}{\cal
R}_{37}{\cal R}_{45}{\cal R}_{46}{\cal R}_{47}{\cal R}_{56}{\cal R}_{67}}
{\mu_1 \ \mu_4 {\cal R}_{14}^2{\cal
R}_{25}^2{\cal R}_{27}^2{\cal R}_{36}^2{\cal R}_{57}^2\prod_{i=1}^{7}{\cal
R}_{ii}}
.}
Here $k^2$ is an integration constant and
\eqn{R}{{\cal R}_{ij}\equiv(\rho^2+\mu_i\mu_j).}
We also define
\eqn{D}{{\cal D}_{ij}\equiv(\mu_i-\mu_j)}
for later convenience. Efficient computation of the conformal factor
requires
a formalism based on going to the complex plane, and is sketched in
Appendix E of \cite{Emparan:2001wk}.

As described near the end of the previous section, we will subtract
solitons and then add them back in, so that (\ref{det}) is automatically
respected. The intuition behind the choice of the seed and
the solitons is based on the analysis of rod-structures \`a la Harmark
(see \cite{Harmark:2004rm}).
In particular, the shapes and locations of the horizons can be
determined from the rod structure, and that gives us a handle on the
geometry without actually trying to analyze the forms of the metric
functions.

The full solution generation process involves the following
steps:

1. Remove an anti-soliton at $a_1$, with trivial BZ vector $(1,0,0)$. This results effectively in
multiplying $(G_0)_{tt}$ by $-\frac{\rho^2}{\mu_1^2}$, upon direct application of (\ref{mu}-\ref{finalmetric}).

2. Remove another anti-soliton at $a_4$, again with trivial BZ vector $(1,0,0)$. This multiplies $(G_0)_{tt}$ by
$-\frac{\rho^2}{\mu_4^2}$.

3. Pull out an overall factor of $-\frac{\rho^4}{\mu_1\mu_4}$ from the
resulting metric. After we are done with the solitonic transformations,
we will put this factor back in. This is a choice of convenience and
nothing prevents us from making that. The resulting metric after these
three steps has the form
\eqn{tildering}{\tilde G_0={\rm
diag}\Big\{\frac{1}{\mu_3\mu_6},
-\frac{\mu_1\mu_3\mu_4\mu_6}{\rho^2\mu_2\mu_5\mu_7},
-\frac{\mu_2\mu_5\mu_7}{\rho^4}\Big\},}
This metric will be our seed for the next transformation, which involves
two solitons. The generating matrix can be computed to be,
\begin{eqnarray}
\label{ringgen}
\tilde\Psi_0={\rm diag}\Big\{
\frac{1}{(\mu_3-\lambda)(\mu_6-\lambda)}, 
\frac{(\mu_1-\lambda)(\mu_3-\lambda)
(\mu_4-\lambda)}{(\mu_2-\lambda)
(\mu_5-\lambda)(\bar\mu_6-\lambda)(\mu_7-\lambda)},
\frac{-(\mu_7-\lambda)}{(\bar\mu_2-\lambda)(\bar\mu_5-\lambda)}
\Big\},\nonumber \\
\end{eqnarray}
where $\bar\mu_i =-\rho^2/\mu_i$.

4. Add two anti-solitons, one at $a_1$ with BZ-vector
$m_0^{(1)}=(1,0,c_1)$ and another at $a_4$, with BZ-vector
$m_0^{(2)}=(1,0,c_2)$, and perform a 2-soliton transformation to obtain
$\tilde G$.

5. Absorb back the factor $-\frac{\rho^4}{\mu_1\mu_4}$ to obtain the final
metric $G$. The conformal factor $f$ can be obtained from
$f_0$ using (\ref{conformal}) with
$\Gamma^{(0)}=\Gamma|_{c_1=c_2=0}$.

Once these transformations are
done, we have the concentric ring solution, except that we
still
need to impose asymptotic flatness and the absence of certain
singularities to make sure that the solution is regular and balanced. We
will address this issue after writing down the
explicit form of the metric. \\

\subsection{The Concentric Ring Solution}

In this section, we write down the functions in the final metric
for the concentric ring
\begin{eqnarray}
\label{ringmetric}
ds^2=G_{tt}dt^2+2G_{t\psi}
dtd\psi
+G_{\psi\psi}
d\psi^2+G_{\phi\phi}d\phi^2+f
(d\rho^2+dz^2), \nonumber
\end{eqnarray}
before imposing regularity etc.
Here, the $\phi\phi$-component is the same as that of the seed metric:
\eqn{phicom}{G_{\phi\phi}=\frac{\mu_3\mu_6\rho^2}{\mu_2\mu_5\mu_7},}
and the conformal factor:
\eqn{confactoring}{f=\frac{
A_1+c_1^2 A_2+2c_1c_2
A_3+c_1^2c_2^2A_4+c_2^2A_5
}{{\cal H}},}
with
\begin{eqnarray}
A_1&=&\mu_2^2 \mu_5^2 \mu_7 {\cal D}_{14}^2 {\cal R}_{12}
{\cal R}_{13}^2{\cal R}_{15}{\cal R}_{16}^2{\cal R}_{17}^2{\cal
R}_{23}{\cal
R}_{24}{\cal R}_{26}{\cal R}_{34}^2{\cal R}_{35}{\cal R}_{37}{\cal
R}_{45}{\cal R}_{46}^2{\cal R}_{47}^2{\cal R}_{56}{\cal R}_{67}, \nonumber
\\
A_2&=&\mu_1^2\mu_2\mu_3\mu_5\mu_6\mu_7^2\rho^2{\cal D}_{12}^2{\cal
D}_{15}^2
{\cal R}_{12}{\cal R}_{14}^2{\cal
R}_{15}{\cal R}_{23}{\cal
R}_{24}{\cal R}_{26}{\cal R}_{34}^2{\cal
R}_{35}{\cal R}_{37}{\cal R}_{45}{\cal R}_{46}^2{\cal
R}_{47}^2{\cal R}_{56}{\cal R}_{67}, \nonumber \\
A_3&=&\mu_1\mu_2\mu_3\mu_4\mu_5\mu_6\mu_7^2 \rho^2{\cal D}_{12}{\cal
D}_{15}{\cal D}_{24}{\cal D}_{45}\times \nonumber \\
&&\hspace{1in}\times{\cal R}_{12}{\cal R}_{14}^2{\cal R}_{15}{\cal
R}_{23}{\cal R}_{24}{\cal R}_{26}{\cal R}_{34}^2{\cal R}_{35}{\cal R}_{37}{\cal
R}_{45}{\cal R}_{46}^2{\cal R}_{47}^2{\cal R}_{56}{\cal R}_{67},\nonumber \\
A_4&=&\rho^{8}\mu_1^2\mu_3^2\mu_4^2\mu_5\mu_6^2\mu_7^3{\cal
D}_{12}^2{\cal
D}_{14}^2{\cal D}_{15}^2{\cal D}_{24}^2{\cal
D}_{45}^2{\cal R}_{12}{\cal R}_{15}{\cal
R}_{23}{\cal R}_{24}{\cal R}_{26}{\cal R}_{35}{\cal
R}_{37}{\cal R}_{45}{\cal R}_{56}{\cal R}_{67}, \nonumber\\
A_5&=&\rho^2\mu_2\mu_3\mu_4^2\mu_5\mu_6\mu_7^2{\cal
D}_{24}^2{\cal
D}_{45}^2{\cal
R}_{12}{\cal
R}_{13}^2{\cal
R}_{14}^2{\cal R}_{15}{\cal
R}_{16}^2{\cal R}_{17}^2{\cal R}_{23}{\cal
R}_{24}{\cal R}_{26}{\cal R}_{35}{\cal R}_{37}{\cal R}_{45}{\cal
R}_{56}{\cal R}_{67},
\nonumber
\end{eqnarray}
and
\eqn{H}{{\cal H}=\mu_1\mu_2\mu_4\mu_5{\cal D}_{14}^2
{\cal R}_{13}{\cal
R}_{14}^2{\cal R}_{16}{\cal
R}_{17}{\cal R}_{25}^2{\cal R}_{27}^2{\cal
R}_{34}{\cal R}_{36}^2{\cal R}_{46}{\cal R}_{47}{\cal
R}_{57}^2\prod_{i=1}^{7}{\cal R}_{ii}.}

The other components of the metric are,
\begin{eqnarray}
G_{tt}= \frac{X_1+c_1^2
X_2+c_2^2X_3+2c_1c_2X_4+c_1^2c_2^2X_5}{\mu_3\mu_6\Delta},\\
G_{t\psi}= \frac{-c_1Y_1
-c_2Y_2+c_1^2c_2Y_3+c_1c_2^2Y_4}{\Delta},\hspace{0.3in}\\
G_{\psi\psi}=\frac{Z_1+c_1^2
Z_2+c_2^2Z_3+2c_1c_2Z_4+c_1^2c_2^2Z_5}{\mu_1\mu_4\Delta},
\end{eqnarray}
with $\Delta\equiv D_1+c_1^2
D_2+c_2^2D_3+2c_1c_2D_4+c_1^2c_2^2D_5$.
The various functions are fully fixed by the following relations,
\begin{eqnarray}
X_1=-\mu_1\mu_4D_1, \ X_2=\rho^2\mu_1\mu_4D_2, \hspace{1.0in} \nonumber \\
X_3=\rho^2\frac{\mu_1}{\mu_4}D_3, \ X_4=\rho^2D_4, \
X_5=-\frac{\rho^4}{\mu_1\mu_4}D_5, \hspace{0.5in} \nonumber \\
Z_1=\mu_2\mu_5\mu_7 D_1, \ Z_2=-\frac{\mu_1^2\mu_2\mu_5\mu_7}{\rho^2}D_2,
\hspace{0.8in}\\
Z_3=-\frac{\mu_2\mu_4^2\mu_5\mu_7}{\rho^2}D_3, \
Z_4=-\frac{\mu_1\mu_2\mu_4\mu_5\mu_7}{\rho^2}D_4, 
Z_5=\frac{\mu_1^2\mu_2\mu_4^2\mu_7}{\mu_5\rho^4}D_5 \hspace{-0.175in}
\nonumber
\end{eqnarray}
and the definitions,
\begin{eqnarray}
D_1&=&\mu_2^2\mu_5^2{\cal D}_{14}^2
{\cal R}_{13}^2{\cal R}_{34}^2{\cal R}_{16}^2{\cal
R}_{46}^2{\cal R}_{17}^2{\cal R}_{47}^2, 
\\
D_2&=&\mu_1^2\mu_2\mu_3\mu_5\mu_6\mu_7\rho^2{\cal D}_{12}^2{\cal D}_{15}^2
{\cal R}_{14}^2{\cal R}_{34}^2{\cal R}_{46}^2{\cal R}_{47}^2,
\\
D_3&=&\mu_2\mu_3\mu_4^2\mu_5\mu_6\mu_7\rho^2{\cal D}_{24}^2{\cal
D}_{45}^2
{\cal R}_{13}^2{\cal R}_{14}^2{\cal R}_{16}^2{\cal
R}_{17}^2,
\\
D_4&=&\mu_1\mu_2\mu_3\mu_4\mu_5\mu_6\mu_7\rho^2{\cal D}_{12}{\cal D}_{24}
{\cal D}_{15}{\cal D}_{45}{\cal R}_{11}{\cal
R}_{44}{\cal R}_{13}{\cal R}_{34}{\cal R}_{16}{\cal R}_{46}{\cal
R}_{17}{\cal R}_{47}, 
\\
D_5&=&\mu_1^2\mu_3^2\mu_4^2\mu_6^2\mu_7^2\rho^8 {\cal D}_{12}^2{\cal
D}_{14}^2{\cal D}_{24}^2{\cal D}_{15}^2
{\cal D}_{45}^2,
\\
Y_1&=&\mu_2^2\mu_5^2\mu_7{\cal D}_{12}{\cal
D}_{14}{\cal D}_{15}
{\cal R}_{11}{\cal R}_{13}{\cal R}_{14}{\cal
R}_{34}{\cal R}_{16}{\cal R}_{46}^2{\cal R}_{17}{\cal R}_{47}^2,
\\
Y_2&=&\mu_2^2\mu_5^2\mu_7{\cal D}_{14}{\cal D}_{24}{\cal
D}_{45}{\cal R}_{44}{\cal R}_{13}^2{\cal
R}_{14}{\cal R}_{34}{\cal R}_{16}^2{\cal R}_{46}{\cal R}_{17}^2{\cal
R}_{47}, 
\\
Y_3&=&\mu_1^2\mu_2\mu_3\mu_5\mu_6\mu_7^2\rho^4{\cal D}_{12}^2{\cal D}_{14}
{\cal D}_{24}{\cal D}_{15}^2{\cal D}_{45}{\cal R}_{44}{\cal
R}_{14}{\cal R}_{34}{\cal R}_{46}{\cal R}_{47}, 
\\
Y_4&=&\mu_2\mu_3\mu_4^2\mu_5\mu_6\mu_7^2\rho^4{\cal D}_{12}{\cal
D}_{14}{\cal D}_{24}^2{\cal D}_{15}{\cal D}_{45}^2{\cal
R}_{11}{\cal R}_{13}{\cal R}_{14}{\cal R}_{16}{\cal R}_{17}. 
 \end{eqnarray}

To complete the solution we need to make sure that it is
asymptotically flat and that there are no conical singularities. 
These conditions will
generate various relations
between the different parameters ($a_i$, $c_1$, $c_2$, 
$k$) in the
solution. This is what we turn to in the next sections.

\subsection{Rod Configuration and Elimination of Singularities}

The rod structure \cite{Harmark:2004rm} for the final solution is useful
for understanding the
horizons, and to see what conditions one has to impose on the
parameters to make sure that there are no singularities. In this
subsection, we discuss the space-like rods because these are the ones that
give rise to the conditions on the parameters. The rod structure is given
in figure 2.
\begin{figure}
\begin{center}
\includegraphics[width=0.9\textwidth]{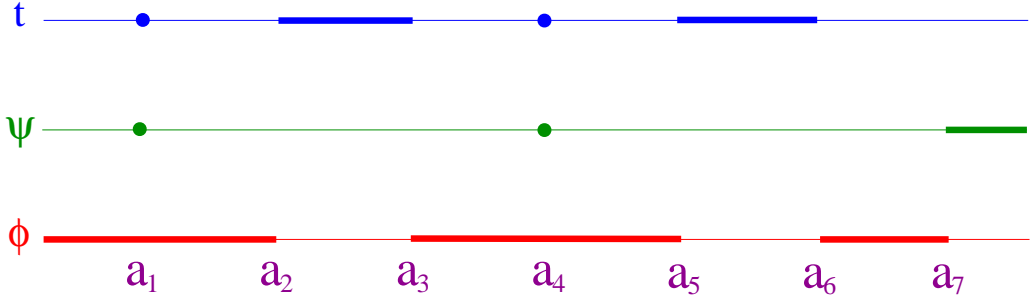}
\caption{Rod configuration for the final solution.}
\label{seedrod}
\end{center}
\end{figure}
The dots in the figure correspond to the locations of the
singularities before we remove them. The final solution can be made
completely regular and then the rod-structure will not have the dots.
\begin{itemize}
\item The semi-infinite rod $(-\infty, \ a_2]$. The direction of the rod,
which is defined as the eigen-direction along which the final metric
matrix $G_{ab}(\rho=0,
z)$ has zero eigenvalues, is $(0,1,0)$. In order to avoid a conical
singularity at the location of the rod, the periodicity of the spacelike
coordinate (here, that would be $\phi$) must be fixed according to the
condition,
\eqn{period}{\Delta \phi =
2\pi\lim_{\rho\rightarrow 0}\sqrt{\frac{\rho^2f}{G_{\phi\phi}}}.}
We end up finding that when $z<a_1$,
\eqn{deltaphi1}{\Delta \phi=2 \pi}
identically, and that when $a_1<z<a_2$,
\eqn{deltaphi12}{\Delta \phi=2
\pi\sqrt{\frac{c_1^2(a_2-a_1)(a_5-a_1)}{2(a_3-a_1)(a_6-a_1)(a_7-a_1)}}= 2\pi.}
Note that the first equality in (\ref{deltaphi12}) is a direct result of imposing (\ref{period}), while the second equality is a condition we are imposing on the parameters so that the period found in (\ref{deltaphi1}) matches that of (\ref{deltaphi12}). The period must be the same for all values of $z$ for the metric to be continuous. 
\item The finite rod $[a_2, \ a_3]$ is timelike and corresponds to the
outer black ring horizon.
\item The finite rod $[a_3, \ a_5]$. The direction of the rod is again
$(0, 1, 0)$, and when $a_3<z<a_4$, to avoid conical singularities we need
\eqn{deltaphi2}{\Delta\phi=2\pi\frac{|Y-Zc_1c_2|}{\sqrt{X}}=2\pi,}
where
\begin{eqnarray}
X&=&\frac{4(a_4-a_1)^2(a_5-a_2)^2(a_6-a_3)^2(a_7-a_2)^2(a_4-a_3)(a_6-a_1)
(a_7-a_1)}{(a_4-a_2)(a_5-a_1)(a_5-a_3)(a_6-a_2)(a_7-a_3)}, \nonumber \\
Y&=&2(a_4-a_3)(a_6-a_1)(a_7-a_1), \nonumber \\
Z&=&(a_2-a_1)(a_5-a_4).\nonumber
\end{eqnarray}
Analogously, when $a_4<z<a_5$, we get
\eqn{deltaphi21}{\Delta\phi=2\pi\frac{|c_1U+c_2V|}{\sqrt{W}}=2\pi,}
with
\begin{eqnarray}
U&=&(a_2-a_1)(a_6-a_4)(a_7-a_4), \nonumber \\
V&=&(a_4-a_2)(a_6-a_1)(a_7-a_1), \nonumber \\
W&=&\frac{2(a_4-a_1)^2(a_5-a_2)^2(a_6-a_1)(a_6-a_3)^2(a_6-a_4)(a_7-a_1)(a_7-a_2)^2(a_7-a_4)}{(a_5-a_1)(a_5-a_4)(a_5-a_3)(a_6-a_2)(a_7-a_3)}.\nonumber
\end{eqnarray}
From both (\ref{deltaphi2}) and (\ref{deltaphi21}), we get constraints on the parameters to avoid conical singularities.
\item The finite rod $[a_5, \ a_6]$ is timelike and corresponds to the
inner black ring horizon.
\item The finite rod $[a_6, \ a_7]$. The direction is $(0, 1,
0)$, and the
periodicity is fixed to be,
\eqn{deltaphi3}{\Delta\phi=2\pi\sqrt{\frac{(a_7-a_1)(a_7-a_4)(a_7-a_3)(a_7-a_6)}{(a_7-a_2)^2(a_7-a_5)^2}}=2\pi.}
\item The semi-infinite rod $[a_7, \ \infty)$. This is the only
$\psi$-rod. The direction is therefore $(0, 0, 1)$. The periodicity is
fixed by a relation analogous to (\ref{period}), and the result is,
\eqn{deltapsi}{\Delta\psi= 2\pi.}
\end{itemize}

One complication that arises in the construction is that there are
singularities in $G_{tt}$ and $G_{\psi\psi}$ that show up at
$z=a_1$ and $z=a_4$. It turns out that we can get rid of these
singularities by setting
\begin{eqnarray}
c_1&=& \sqrt{\frac{2(a_3-a_1)(a_6-a_1)(a_7-a_1)}{(a_2-a_1)(a_5-a_1)}}, \label{1}
\\
c_2&=& \sqrt{\frac{2(a_4-a_3)(a_6-a_4)(a_7-a_4)}{(a_4-a_2)(a_5-a_4)}}. \label{2}
\end{eqnarray}
It should be noted that the first of these conditions is identical to the condition that fixes the
periodicity of the
$\phi$ rod at $[a_1, a_2]$  to $2\pi$, because of (\ref{deltaphi12}). There is an ambiguity in the choice of the sign of each $c_i$. This is physical: we will see later that it is related to the direction of rotation of each ring. For most of what follows we will assume for definiteness that both $c_i$ are positive, but things go through essentially unchanged for other choices of sign, except for a minor caveat we will emphasize when we compute the ADM quantities. Note also that the $a_i$ are dimensionful, but the conical deficit angles we have calculated are dimensionless as they should be. 

The fact that the horizon is two disconnected rings, is also evident from the rod diagram. The way to see this is to note that there are no time-like rods adjacent to the semi-infinite $\psi$-rod. If one treats the tip of the $\psi$-rod ($a_7$) as the origin of $z$-axis (which is allowed because of translational invariance along $z$), then this means that the horizon starts away from the center of the geometry. Together with the fact that the system is axi-symmetric, similar arguments immediately lead us to the conclusion that there are two ring-shaped horizons, and that they are concentric. Our rod diagram can be compared to the rod diagrams for flat Minkowski space, Myers-Perry black hole, the black Ring and the black Saturn, and they all fit together neatly.

\label{as}
\subsection{Asymptotic Flatness}

It is possible to verify \cite{Harmark:2004rm} that the asymptotic region
is given by the conditions,
\eqn{asymp}{ \sqrt{\rho^2 + z^2} \rightarrow \infty, \ \ \ {\rm with}
 \ \frac{z}{\sqrt{\rho^2
+ z^2}} \ \ \ \ {\rm finite}.}
Introducing coordinates $r$ and $\theta$ according to \cite{Elvang:2007rd}
\eqn{rtheta}{\rho=\frac{1}{2}r^2\sin 2\theta, \ \
z=\frac{1}{2}r^2\cos2\theta,}
the asymptotic limit is succinctly contained in $r \rightarrow \infty$.
At infinity, we want the black di-ring metric to reduce to the form
\eqn{asymp}{ds^2=-dt^2+dr^2+r^2 d\theta^2+r^2\sin^2\theta
d\psi^2+r^2\cos^2\theta d\phi^2.}
It is possible to check that $G_{tt}, \ G_{t\psi}, \ G_{\psi\psi}, \
G_{\phi\phi}$ go to the right limits as $r \rightarrow \infty$.

So far we have left the integration constant $k^2$ in the conformal factor
$f$ to
be arbitrary. In fact the condition that
\eqn{fasmyp}{f (d\rho^2+dz^2) \rightarrow dr^2+r^2d\theta^2,}
at infinity fixes $k^2=1$.

Once we impose all these conditions, we have a fully regular and balanced asymptotically flat di-ring. It is encouraging that the asymptotic flatness conditions do not result in too many further conditions on the parameters, for physical reasons. We elaborate on the counting of parameters in the next subsection.

\subsection{Parameter Counting}

Lets count the number of parameters of the black di-ring. We have 7 parameters $a_i$, two BZ parameters $c_i$, and $k^2$. So in total we had 10 parameters to begin with. Only the relative positions of the $a_i$ matter because of translational invariance along $z$, so we can define \cite{Elvang:2007rd}
\bea
L^2=(a_7-a_1)
\eea
as a convenient length scale. Together with $a_2,...a_6$, this leaves us with six variables. We saw above that $k^2$ is set to 1. The $c_i$ are determined by $a_i$ according to (\ref{1}) and (\ref{2}), so we are still left with six independent parameters at this stage. The conical singularity constraints from last section give rise to three more independent constraints, and so finally we end up with 3 independent parameters for the black di-ring.

Happily, this is what one would expect on general grounds. The di-ring should have two indpendent masses and two-independent angular momenta, one each for each of the two rings. But one of these four can be scaled away because classical gravity is a conformal theory, see  \cite{EEF} for a nice discussion of this. So indeed we expect to have three independent parameters.

It should be noted that the three non-trivial constraints arising from the absence of conical singularities, fix the parameters only implicitly. We have not been able to solve them analytically in a useful way. But the use of the metric, especially in investigations of thermodynamical phases etc., is bound to be numerical, so this is not a serious problem. In particular, the fact that the constraints on the parameters is implicit, should {\em not} be taken to mean that the constraints are inconsistent. The most direct way to demonstrate this is to find explicit values for the $a_i$ which satisfy the constraints. To do this, first introduce the variables $z_i$ which are defined as
\bea
z_i=\frac{a_{i+1}-a_1}{L^2}.
\eea
Notice that the sequence $0, z_1, z_2, z_3,z_4, z_5, 1$ is non-decreasing. We can rewrite the conical singularity elimination conditions of the last subsection in terms of these new variables. The advantage is that the overall scale $L$ drops off from all expressions, so we only have to deal with\footnote{Note that we also have to use the expressions (\ref{1}) and (\ref{2}) to solve for $c_i$.} the $z_i$. Now, we are left with three equations and five variables, and our aim is to show that there are no inconsistencies.

Generically of course, such a system is well-posed, our aim is to merely make sure that what we have is not some degenerate, inconsistent special case. This is easy to do numerically 
by starting with seeds for two of the $z_i$ and solving for the remaining three using the constraint equations. The result is a consistent solution if and only if the resulting $z_i$ satisfy the non-decreasing property. When we do this, we find that there are indeed solutions. We present an example with the seed $z_1=0.3, z_2=0.4$, below:
\bea
\{z_1=0.3,z_2=0.4, z_3=0.678153, z_4=0.743009, z_5=0.832417\}.
\eea
It can be checked by direct substitution that these values solve the constraint equations (the scale $L$ does not affect this). More solutions can be found by a numerical scanning starting from this seed. A more exhaustive scanning strategy would be to systematically scan for $z_i$ between 0 and 1 using some appropriate bin-size. Finding all interesting solutions is likely to require an adaptive bin-size scanning strategy, because we don't know the measure on the moduli space of the $z_i$: in particular, it can have structures at various resolutions depending on where we are. A similar situation was encountered in \cite{metastable} for the black Saturn as well. We strongly suspect that the space of solutions densely fills out at least part of the phase space considered in \cite{EEF}, but we leave the details for future work. It would also be interesting to see which of these phases go away, when we impose thermodynamic equilibrium between the two rings.

A more analytical, but less concrete, piece of evidence for existence of solutions is that there exist limits where we can reduce the solution to the single ring form. The fact that the well-known ring solution can be found in the boundary of the moduli space of our di-ring solutions is another indication that the moduli space is non-vacuous.
Indeed, we can obtain the black ring of Emparan and Reall as a limit of our di-ring solution. A hint on how to do this can be found by comparing our final rod diagram with the black ring rod diagram \cite{Harmark:2004rm}: we set $a_2=a_3$ and $a_1=a_3$. After some massaging, the metric functions can be brought to the form of the black ring metric as written in the coordinates presented in (A.7-A.10) in \cite{Elvang:2007rd}, if we do the following replacements: $c_1\rightarrow c_2$ with the other subscripts renamed as $1 \rightarrow 4, 3 \rightarrow 7, 4 \rightarrow 6, 5 \rightarrow 5$.
Here the left-hand sides correspond to the notations in \cite{Elvang:2007rd} and the right hand sides corresponds to our notations. The singularity removal conditions also reduce to the corresponding conditions for the black ring.

We got the single ring in the above limit by (effectively) removing the outer black ring. An exactly analogous construction can be done by removing the inner black ring. We have checked that this also results in a single black ring solution as expected.

\subsection{ADM Mass and Angular Momentum}

The ADM mass and angular momentum of the solution can be computed using 
the metric functions, extending our results on asymptotic flatness. The basic idea is to expand the metric functions in the coordinates defined in section \ref{as}, and to identify the mass and angular momentum from the fall-offs, see section 4.3 of \cite{Harmark:2004rm}. One way to simplify the computation is to go to infinity along the direction $\theta=\frac{\pi}{4}$ so that we can set $z=0$. Keeping track of the leading and sub-leading terms, once the dust settles we end up with
\bea
GM_{ADM}= \frac{3\pi }{4} \times 
\Big( {a_6}-{a_4}+{a_3}-{a_1}\Big)
, \hspace{1.2in}\nonumber \\
GJ_{ADM}=\pi \frac{(a_2 - a_1) (a_5 - a_1) c_1 + (a_4 - a_2) (a_5 - a_4) c_2 }{
  2 (a_4 - a_1)} \hspace{0.7in}
\eea
Considering the formidable form of the di-ring metric, one might get the impression that these expressions are rather simple. But one should remember that the conical deficit constraints and the relations relating $c_i$ to $a_i$ are yet to be applied to these relations, and this can only be done numerically. In this sense, the di-ring solution is more complicated than the Saturn solution.

It is intuitively clear from the expression for $J$ that the choice of sign of $c_i$ is directly related to the direction of rotation of each ring. In terms of the scale $L$ that we introduced, $G M \sim L^2$, while $G J \sim L^3$, which is expected both from general principles and also from the specific expressions obtained previously in the literature e.g., for the case of the black Saturn.

It is also worth mentioning that the ADM mass presented above is manifestly positive (as it should be) as an automatic consequence of the ordering of the solitons.







\section{\bf Discussion}

The purpose of this paper was to present a derivation of the black
di-ring using the inverse scattering method. In this concluding section,
we make some comments about our approach and about the di-ring solution.

The implementation of the inverse scattering method adopted here for the
construction of the di-ring differs from
the approach used in \cite{Tomizawa:2005wv} for the construction of some
other axially symmetric solutions. There the condition on the
determinant (\ref{det}) was imposed by demanding that the solitonic
transformations be limited to a 2 $\times$ 2 block, and then renormalizing
(\ref{detG}) appropriately. Instead, we keep the transformations general,
following the idea presented in \cite{Pomeransky:2005sj, Elvang:2007rd}.
The advantage of
this approach is
that it is sufficiently general to allow the possibility of
constructing
more complicated axi-symmetric vacuum solutions: we hope to
return to the construction of some of
these generalizations in the future.

The black di-ring solution that we found is somewhat more complicated in its final form than
the Saturn solution. This is expected, since the hole there is replaced here with another ring,
and the latter is a more complicated object. Still, we found that the solution can be brought to a form
that is numerically as tractable as the Saturn itself\footnote{This is not to say that either of these solutions is easy to explore, even numerically!}. This opens up the possibility of exploring questions regarding higher dimensional black holes in the context of the black di-ring.  One could also investigate the
physics and thermodynamics of the di-ring solution. Similar analyses have been done for the black Saturn, where effects like frame-dragging were explicitly checked. It would be interesting to see if there exists a parameter range where the two rings in our solution can be in thermodynamic equilibrium, see \cite{EEF, metastable}. Related questions are under investigation.

\section{Acknowledgments}

It is a pleasure to thank
Henriette Elvang for useful conversations and correspondence.  We would
also like to thank Carlo Maccaferri, Stanislav
Kuperstein and Daniel Persson for inspiration.
This work is supported in part by IISN - Belgium (convention
4.4505.86), by the Belgian National
Lottery, by the
European Commission FP6 RTN programme MRTN-CT-2004-005104 in which the
authors
are associated with V. U. Brussel, and by the Belgian Federal Science
Policy Office through the Interuniversity Attraction Pole P5/27.



\newpage

\end{document}